\documentclass[aps,twocolumn,superscriptaddress,showpacs,letter]{revtex4-1}
\usepackage{graphicx}
\usepackage{dcolumn}
\usepackage{bm}
\usepackage{cancel}
\usepackage{color}             
\usepackage{epsfig}
\usepackage{amssymb}
\usepackage{amsmath}

\begin{document}
      
\title{QCD effective charges and the structure function $F_{2}$ at small-$x$: Higher twist effects}

\author{D.~Hadjimichef}
\email{dimiter@if.ufrgs.br}
\affiliation{Instituto de F\'isica, Universidade Federal do Rio Grande do Sul, Caixa Postal 15051, 91501-970, Porto Alegre, RS, Brazil}
\author{E.~G.~S.~Luna}
\email{luna@if.ufrgs.br}
\affiliation{Instituto de F\'isica, Universidade Federal do Rio Grande do Sul, Caixa Postal 15051, 91501-970, Porto Alegre, RS, Brazil}
\affiliation{Instituto de F\'{\i}sica, Facultad de Ingenier\'{\i}a, Universidad de la Rep\'ublica, \\
  J.H. y Reissig 565, 11000 Montevideo, Uruguay}
\author{M.~Pel\'aez}
\email{mpelaez@fing.edu.uy}
\affiliation{Instituto de F\'{\i}sica, Facultad de Ingenier\'{\i}a, Universidad de la Rep\'ublica, \\
J.H. y Reissig 565, 11000 Montevideo, Uruguay}

\begin{abstract}

We consider the effect of higher twist operators of the Wilson operator product expansion in the structure function $F_{2}(x,Q^{2})$ at small-$x$,
taking into account QCD effective charges whose infrared behavior is constrained by a dynamical mass scale.
The higher twist corrections are obtained from the renormalon formalism.
Our analysis is performed within the conventional framework of next-to-leading order, with the factorization and renormalization scales chosen to
be $Q^{2}$.
The infrared properties of QCD are treated in the context of the generalized double-asymptotic-scaling approximation. 
We show that the corrections to $F_{2}$ associated with twist-four and twist-six are both necessary and sufficient for a good description of the
deep infrared experimental data.

\end{abstract}

\maketitle

\section{Introduction}

In the description of the structure function $F_{2}(x,Q^{2})$ of the proton, a long-standing question is the extent to which the nonperturbative
properties of QCD affect the behavior of $F_{2}$.
The necessity of nonperturbative corrections arises as follows: at sufficiently small $x$ the power series in $\alpha_s \ln(1/x)$ may be
resummed via BFKL equation \cite{bfkl}. The result of resumming
these leading terms is sensitive to the infrared $k_{T}$ region and, for running $\alpha_s$, it is found that \cite{martin01}
\begin{eqnarray}
\tilde{g}(x,k_{T}^2) \sim C(k_{T}^2)\, x^{-\lambda} \, ,
\label{g1}
\end{eqnarray}
where $\lambda \sim 0.5$ and $\tilde{g}(x,k_{T}^{2})$ is here the {\sl unintegrated} gluon distribution. The relation between
$\tilde{g}(x,k_{T}^{2})$ and $g(x,Q^{2})$, the {\sl standard} gluon distribution to be determined by the analysis of the $F_{2}(x,Q^{2})$ data, reads
\begin{eqnarray}
\tilde{g}(x,k_{T}^2) = \left. \frac{\partial (xg(x,Q^{2}))}{\partial \ln Q^{2}} \right|_{Q^{2}=k_{T}^2} .
\end{eqnarray}
At this point it is clear that nonperturbative contributions are needed. First, the resummation program requires knowledge of the gluon for all
$k_{T}^2$ including the deep infrared region. Unfortunately, in this confinement region the BFKL equation is not expected to be valid.
Second, the HERA data on $F_{2}$ imply a steep gluon at small-$x$ and, in addition, such a steep behavior can be generated from a flat-$x$ gluon
distribution at some initial low $Q_{0}^{2}$ scale, i.e. the data in the small-$x$ region show that $F_{2}$ tend to a flat shape with
decreasing $Q^{2}$, particularly for low $Q^{2}$. This indicates that the singular behavior $x^{-\lambda}$ predicted by BFKL must be suppressed by
nonperturbative effects.
Hence approaching the low $Q^{2}$ region from the QCD theory makes evident the problem of how to incorporate in an effective way nonperturbative
corrections into the description of the structure function $F_{2}$.

Fortunately, this problem was properly addressed some time ago \cite{luna001} by considering
the possibility that the nonperturbative dynamics of QCD generate a dynamical gluon mass at very slow $Q^{2}$
region. This effective mass is intrinsically related to a finite strong coupling constant, and its existence is strongly supported
either by QCD lattice simulations \cite{lattice002,lattice001,othergauge001} or by phenomenological results \cite{halzen,luna002}.
In \cite{luna001} the investigation was focused exclusively on the small-$x$ region since this condition establish
the main criteria for the validity of the so called generalized double-asymptotic-scaling (GDAS) approximation \cite{kot,kot2,cve},
which in turn is particularly relevant to the analysis since it is consistent with the phenomenon of dynamical mass generation in
QCD \cite{luna001,luna003}. The importance of the GDAS approximation can be understood as follow. By analyzing exclusively the
small-$x$ region, some of the simpler existing analytical solutions of the DGLAP evolution equation \cite{dglap} in the small-$x$ limit can be
directly used \cite{ball001,ball002,frichter,yund1,kot3}. Within this approach the $F_{2}$ data at small-$x$ can be interpreted in terms of the
double-asymptotic-scaling (DAS) phenomenon \cite{ball001,ball002}, where small-$x$ nucleon structure functions exhibit scaling in
two new variables, provided only that the small-$x$ behavior of the parton distribution functions (PDFs) at some starting point $Q^{2}_{0}$ is
sufficiently soft.
The resulting analytical solutions can in turn be extended in order to
include the subasymptotic part of the DGLAP evolution \cite{kot,kot2,ball001,mank}, in what is finally called GDAS approximation, leading
to the prediction of flat forms at small-$x$ for parton distributions at some
input scale $Q^{2}_{0}$, namely
\begin{eqnarray}
f_a (x,Q^2_{0}) = A_a  \,\,\,\,\,\,\,\,\, (a=q,g)  \,\, ,
\label{eq1}
\end{eqnarray}
where $A_a$ are unknown parameters to be determined from the data. Note that in the above expression the parton distribution functions
are multiplied each by $x$, that is $f_{g}(x,Q^2)\equiv xg(x,Q^2)$, $f_{q}(x,Q^2)\equiv xq(x,Q^2)$. In summary, in Ref. \cite{luna001} the structure
function $F_{2}(x,Q^2)$ was described by means of the GDAS approximation, exploring the close interrelation between the flat behavior of initial
parton distributions and the phenomenon of dynamical mass generation in QCD.

In this Letter we extend the previous study in three significant ways. 
First, we have included deep infrared $F_{2}$ data in our global fits. The new dataset includes measurements at $Q^2 = 0.2$ GeV$^2$ and
$Q^2 = 0.25$ GeV$^2$. The inclusion of deep infrared data is observed to have quite a large impact on the overall result. 
Second, we have introduced higher twist corrections to $F_{2}$ for the case, studied in the previous work, of a flat initial condition for the
leading twist QCD evolution. These higher twist corrections are estimated from the infrared renormalon model \cite{beneke001,stein001}.
Third, we have investigated, in addition to the logarithmic and power-law cases, a QCD effective charge built from a particular case of the
Curci-Ferrari Lagrangians.

The Letter is organized as follows: in the next section we introduce higher twist contributions to $F_{2}$ from the renormalon formalism and
present the formalism for analyzing the structure function of the proton in the GDAS approximation.
In the Sec. III we discuss the definition of the QCD effective charge and present three versions of holomorphic charges.
Our results are presented in the Sec. IV, where we compare our results on the structure function $F_{2}$ with experimental data. Our analysis is
carried on using the formalism developed in the sections II and III. In Sec. V we draw our conclusions and final remarks.

\section{Higher twist operators}

Higher twist corrections to deep inelastic scattering (DIS) processes have been studied systematically in the framework of the
operator product expansion (OPE) \cite{dis007}. It has to be admitted at once that the details of this corrections are not yet fully understood,
owing to the theoretical difficulties of controlling power corrections in effective theories \cite{martinelli001}: the calculation of power
corrections requires the evaluation of the matrix elements of higher-twist operators, but in order to cancel renormalon ambiguities it is
also necessary to compute the Wilson coefficient functions to sufficiently high orders of
the perturbation series. These renormalon ambiguities are of the same order as the power corrections.

In the case of twist-four operators, for example, radiative corrections generate terms of the form of a square of the (dimensionful) ultraviolet
cut-off multiplied by the
lower-order (dimensionless) matrix element of the twist-two. These mixing terms make the definition of twist-four contributions ambiguous.
However, in the operator product expansion of DIS structure functions, this quadratic ambiguity always cancels against the corresponding
ambiguity in the definition of the twist-two contribution. As a result, if the twist-two and twist-four contributions are calculated within the
same regularization scheme, the sum of both contributions is unambiguous up to order $1/Q^2$ \cite{beneke001,stein001}.

Unfortunately, given that in general only a few terms of the perturbative series are known, and furthermore these series are plagued by similar
renormalon ambiguities, it is not clear if the ambiguity of higher twist contributions can also be canceled \cite{martinelli001}. However, the
subtle relation between 
the twist-two and the twist-four contributions has inspired the hypothesis that the main contributions to the matrix elements of the twist-four
operators are proportional to their (quadratically) divergent parts \cite{beneke001}. This means that in practice we can obtain information
about power corrections from the large-order behavior of the corresponding series. This approach is called infrared renormalon model.

We shall from now on present the higher-twist contributions to the DIS structure function $F_{2}(x,Q^{2})$ in the framework of the renormalon
formalism, as calculated by Illarionov, Kotikov and Parente in \cite{kot2}.
We start by writing down the twist-two term of $F_{2}(x,Q^{2})$ at NLO \cite{kot,kot2,mank}:
\begin{eqnarray}
\frac{1}{e} F^{\tau2}_{2}(x, Q^{2}) =  f^{\tau2}_{q}(x, Q^{2})  +  \frac{4T_{R}n_{f}}{3}\,
\frac{\alpha_{s}(Q^{2})}{4\pi} f^{\tau2}_{g}(x, Q^{2}), \nonumber \\
\label{f211}
\end{eqnarray}
where $n_{f}$ is the effective number of quarks, 
$e=\sum_{i}^{f} e_{i}^{2}/n_{f}$ is the average charge squared,
$T_{R}=1/2$ is the color factor for $g\to q\bar{q}$ splitting. It may be worth emphasizing that the expression (\ref{f211}) is valid only for
$x \ll 1$. Remember that for all range of $x$ the NLO term of $F_{2}(x,Q^{2})$ is, in fact, given
by
\begin{eqnarray}
\frac{1}{e} F^{\tau2}_{2}(x, Q^{2}) = \sum_{a=q,g} \left[ B_{2,a}(x) \otimes f_{a}(x,Q^{2})  \right],
\label{f211bf}
\end{eqnarray}
where
\begin{eqnarray}
B_{2,q}(x) = \delta(1 - x) + \frac{\alpha_{s}}{4\pi}\, B^{(1)}_{2,q}(x),
\end{eqnarray}
\begin{eqnarray}
B_{2,g}(x) = \frac{\alpha_{s}}{4\pi}\, B^{(1)}_{2,q}(x);
\end{eqnarray}
in the above expressions the superscript $(1)$ indicates next-to-leading order, and  the symbol $\otimes$ stands for the convolution
formula
\begin{eqnarray}
A(x) \otimes B(x) = \int_{x}^{1} \frac{dy}{y}\, A(y) B\left( \frac{x}{y} \right);
\end{eqnarray}
more details about the computation of the coefficient functions $B_{2,a}(x)$ and the expression (\ref{f211}) may be found in the appendix
A of \cite{kot2}.

Returning now to $F^{\tau2}_{2}(x, Q^{2})$, the parton distributions can be written as
\begin{eqnarray}
f^{\tau2}_{a}(x,Q^{2}) = f_{a}^{\tau2,+}(x,Q^{2}) + f_{a}^{\tau2,-}(x,Q^{2}),
\label{f222}
\end{eqnarray}
with $a=q,g$. Here the ``$+$'' and ``$-$'' representation follows from the solution, at twist-two approximation, of the DGLAP equation
in the Mellin moment space \cite{kot}:
\begin{eqnarray}
f_{a}^{\tau2,-}(x, Q^{2}) &=& A_{a}^{-}(Q^{2},Q_{0}^{2}) \nonumber \\
 & & \exp \left[ -d_{-}(1)s-D_{-}(1)p \right] + {\cal O}(x),
\label{f233}
\end{eqnarray}
\begin{eqnarray}
f_{g}^{\tau2,+}(x, Q^{2}) &=& A_{g}^{+}(Q^{2},Q_{0}^{2}) \tilde{I}_{0}(\sigma) \nonumber \\
 & & \exp \left[ -\bar{d}_{+}(1)s-\bar{D}_{+}(1) p \right] + {\cal O}(\rho),
\label{f255}
\end{eqnarray}
\begin{eqnarray}
f_{q}^{\tau2,+}(x, Q^{2}) &=& A_{q}^{+}(Q^{2},Q_{0}^{2}) \left[ \left( 1-\bar{d}_{+-}^{q}(1)\,
\frac{\alpha_{s}(Q^{2})}{4\pi} \right) \right. \nonumber \\
& & \times \left. \rho \tilde{I}_{1}(\sigma) + \frac{20 C_{A}}{3}\, \frac{\alpha_{s}(Q^{2})}{4\pi}
\, \tilde{I}_{0}(\sigma) \right] \nonumber \\
& & \times \exp \left[ -\bar{d}_{+}(1)s - \bar{D}_{+}(1)p \right] + {\cal O}(\rho)\!,
\label{f244}
\end{eqnarray}
where 
\begin{eqnarray}
p = \frac{1}{4\pi} \left[ \alpha_s(Q^2_0 ) - \alpha_s(Q^2) \right] ,
\end{eqnarray}
\begin{eqnarray}
s = \ln \left[ \alpha_s(Q^2_0 )/\alpha_s(Q^2) \right],
\end{eqnarray}
\begin{eqnarray}
D_\pm (n)= d_{\pm\pm} (n)-(\beta_1 /\beta_0) d_\pm (n),
\end{eqnarray}
\begin{eqnarray}
\sigma = 2\sqrt{\left( \hat{d}_{+}s+\hat{D}_{+}p \right) \ln x}
\end{eqnarray}
and
\begin{eqnarray}
\rho = \sqrt{ \left( \hat{d}_{+}s+\hat{D}_{+}p \right) /\ln x}
= \frac{\sigma}{2\ln (1/x)};
\end{eqnarray}
in the above expressions $\beta_{0}$ $(\beta_{1})$ is the first (second) coefficient of the $\beta$ function of the QCD.  
The components of the anomalous dimension $d_{-}(n)$ and of the regular ($\bar{d}$) and singular ($\hat{d}$) parts of
$d_{+}(n) = \hat{d}_{+}/(n-1) + \bar{d}_{+}(n)$, in the limit $n\to 1$, are given by
\begin{eqnarray}
\hat{d}_{+} = -\frac{4C_{A}}{\beta_{0}},
\end{eqnarray}
\begin{eqnarray}
\bar{d}_{+}(1) = 1 + \frac{8T_{R}n_{f}}{3\beta_{0}} \left( 1 - \frac{C_{F}}{C_{A}} \right)
\end{eqnarray}
and
\begin{eqnarray}
d_{-}(1) = \frac{8T_{R}C_{F}n_{f}}{3C_{A}\beta_{0}}.
\end{eqnarray}
The functions $\tilde{I}_{\nu}$ ($\nu = 0,1$) are functions related to the Bessel function $J_{\nu}$ and the modified Bessel function $I_{\nu}$ by
\begin{eqnarray}
\tilde{I}_{\nu}(\sigma) = \left\{ \begin{array}{cll}
 i^{\nu}J_{\nu}(\bar{\sigma}), &   \mbox{if } \sigma^{2} =  -\bar{\sigma}^{2} < 0, & \\
 I_{\nu}(\bar{\sigma}), & \ \mbox{if } \sigma^{2} =  \bar{\sigma}^{2} \geq 0,
 \end{array}\right.  \nonumber
\end{eqnarray}
and the factors $A_{a}^{+,-}$, as well as the components of the regular and singular parts of the anomalous dimensions
$D_\pm$ are given by
\begin{eqnarray}
A_{g}^{+}(Q^{2},Q_{0}^{2}) &=& \left[ 1 - \bar{d}_{+-}^{g}(1) \,
  \frac{\alpha_{s}(Q^{2})}{4\pi} \right] A_{g} \nonumber \\
 & & + \frac{C_{F}}{C_{A}}\left[ 1 - d_{-+}^{g}(1)\,
  \frac{\alpha_{s}(Q_{0}^{2})}{4\pi} \right. \nonumber \\
 & & - \left. \bar{d}_{+-}^{g}(1) \, \frac{\alpha_{s}(Q^{2})}{4\pi} \right] A_{q} \, ,
\end{eqnarray}
\begin{eqnarray}
A_{g}^{-}(Q^{2},Q_{0}^{2}) = A_{g} - A_{g}^{+}(Q^{2},Q_{0}^{2}) \, ,
\end{eqnarray}
\begin{eqnarray}
A_{q}^{+}(Q^{2},Q_{0}^{2}) = \frac{2T_{R}n_{f}}{3C_{A}} \left( A_{g} +
\frac{C_{F}}{C_{A}}\, A_{q} \right) \, ,
\end{eqnarray}
\begin{eqnarray}
A_{q}^{-} (Q^{2},Q_{0}^{2}) = A_{q}- \frac{20C_{A}}{3}\,
\frac{\alpha_{s}(Q_{0}^{2})}{4\pi} \, A_{q}^{+}(Q^{2},Q_{0}^{2}) \, ,
\end{eqnarray}
\begin{eqnarray}
\hat{d}_{++} = \frac{8T_{R}n_{f}}{9\beta_{0}} \left( 23C_{A}-26C_{F}
\right),
\end{eqnarray}
\begin{eqnarray}
\hat{d}_{+-}^{q} = -\frac{20C_{A}}{3} , \hspace{1.3truecm}
\hat{d}_{+-}^{g} = 0 ,
\end{eqnarray}
\begin{eqnarray}
\bar{d}_{++}(1) &=& \frac{8}{3\beta_{0}}\left[
  \frac{C_{A}^{2}}{3}\left(36\zeta(3)+33\zeta(2)
- \frac{1643}{12}\right) \right. \nonumber \\
 & & - \left(2C_{F}\zeta(2)+\frac{43}{9} \, C_{A}-\frac{547}{36} \,
C_{F} + \frac{3}{2}\frac{C_{F}^{2}}{C_{A}}\right)n_{f} \nonumber \\
& & - \left. \frac{13}{18} \frac{C_{F}}{C_{A}} \left(1-2 \,
\frac{C_{F}}{C_{A}}\right)n_{f}^{2}\right] ,
\end{eqnarray}
\begin{eqnarray}
d_{--}(1) &=& \frac{4C_{A}C_{F}}{\beta_{0}}\left(1 - 2 \,
\frac{C_{F}}{C_{A}}\right) \left[ 2\zeta(3)-3\zeta(2) \right. \nonumber \\
 & & + \left. \frac{13}{4} + \frac{13}{27}\frac{n_{f}^{2}}{C_{A}^{2}} \right] \nonumber \\
 & & + \frac{4C_{F}}{3\beta_{0}}\left(4\zeta(2)-\frac{47}{18} + 3 \,
\frac{C_{F}}{C_{A}}\right)n_{f},
\end{eqnarray}
\begin{eqnarray}
\bar{d}_{+-}^{q}(1) &=& C_{A}\left(9-3 \, \frac{C_{F}}{C_{A}}-4\zeta(2)\right) \nonumber \\
 & & - \frac{13}{9}\left(1-2
\, \frac{C_{F}}{C_{A}}\right)n_{f} \, ,
\end{eqnarray}
\begin{eqnarray}
\bar{d}_{+-}^{g}(1) = \frac{40T_{R}n_{f}}{9} \frac{C_{F}}{C_{A}} , 
\end{eqnarray}
\begin{eqnarray}
d_{-+}^{q}(1) = 0 
\end{eqnarray}
and
\begin{eqnarray}
d_{-+}^{g}(1) = -\left[C_{A}+\frac{2}{3}\left(1-2\frac{C_{F}}{C_{A}}\right)T_{R}n_{f}\right] ,
\end{eqnarray}
where $\zeta$ is the Riemann zeta function, and $C_{A}=N$ and $C_{F}=(N^{2}-1)/2N$ are Casimir color-factors, $N^{2}-1$ being the dimension of
the group $SU(N)$.

Higher twist estimations (twist-four and twist-six) are known, in the framework of the infrared renormalon formalism, for the nonsinglet case
\cite{dasgupta001} as well to the singlet one \cite{stein001}. In this approach the higher twist corrections can be expressed in terms of the
leading twist expansion of $F_{2}(x,Q^{2})$. As previously indicated, we adopt the formalism and notation put forwarded in \cite{kot2}. 
In this case the twist-four ($\tau 4$) correction to $F_{2}(x,Q^{2})$ in the $[R]$enormalon formalism is given by

\begin{eqnarray}
F^{[R]\tau 4}_{2}(x,Q^{2}) &=& e \sum_{a=q,g} A^{\tau 4}_{a} \tilde{\mu}^{\tau 4}_{a} (x,Q^{2}) \otimes f^{\tau 2}_{a} (x,Q^{2}) \nonumber \\
 &=& \sum_{a=q,g} F^{[R]\tau 4}_{2,a} (x,Q^{2}) ,
\label{renormalont4}
\end{eqnarray}
where $f^{\tau 2}_{a} (x,Q^{2})$ are the gluon and quark-singlet PDFs in twist-two ($\tau 2$) approximation, whereas
$\tilde{\mu}^{\tau 4}_{a}(x,Q^{2})$
are just the functions obtained in \cite{stein001} by means of the infrared renormalon model. Therefore, once we have found the twist-four
correction, we can write the structure function as

\begin{eqnarray}
F^{[R]}_{2}(x,Q^{2}) = F^{\tau 2}_{2}(x,Q^{2}) + \frac{1}{Q^{2}}\, F^{[R]\tau 4}_{2}(x,Q^{2}).
\end{eqnarray}

In Eq. (\ref{renormalont4}) the Mellin transform of $\tilde{\mu}^{\tau 4}_{a}(x,Q^{2})$ is given by
\begin{eqnarray}
\mu^{\tau 4}_{a} (n,Q^{2}) = \int_{0}^{1} dx \, x^{n-1} \, \tilde{\mu}^{\tau 4}_{a} (x,Q^{2}) .
\end{eqnarray}

As in the case of the twist-two expansion, $F^{[R]\tau 4}_{2}(x,Q^{2})$ can also be split into two parts, namely the ``+'' and the ``-''
representations:
\begin{eqnarray}
F^{[R]\tau 4}_{2}(x,Q^{2}) = F^{[R]\tau 4,+}_{2}(x,Q^{2}) + F^{[R]\tau 4,-}_{2}(x,Q^{2}). \nonumber \\
\end{eqnarray}

Thus, applying the technique of transforming the Mellin convolutions (taking the limit $n \to 1$) to simple products at small $x$
\cite{kotikov009} (these products replace the convolution of two functions at small $x$), the ``+'' and the ``-'' parts of
$F^{[R]\tau 4}_{2}(x,Q^{2})$ can be written as \cite{kot2}:
\begin{widetext}
\begin{eqnarray}
\frac{1}{e} \, F^{[R]\tau 4,+}_{2}(x,Q^{2}) &=& \frac{32 T_{R} n_{f}}{15 \beta_{0}^{2}} \, f^{\tau 2,+}_{g}(x,Q^{2})
\left\{ A^{\tau 4}_{g} \left( \frac{2}{\rho} \frac{\tilde{I}_{1}(\rho)}{\tilde{I}_{0}(\rho)}
+ \ln \left( \frac{Q^{2}}{|A^{\tau 4}_{g}|} \right) \right) + \frac{4 C_{F} T_{R} n_{f}}{3 C_{A}} \, A^{\tau 4}_{q} 
\left[ \left( 1 - \bar{d}^{q}_{+-}(1)a_{s}(Q^{2}) \right) \right. \right. \nonumber \\
 & & \times \left. \left. \left( \frac{2}{\rho} \frac{\tilde{I}_{1}(\rho)}{\tilde{I}_{0}(\rho)}
+ \ln \left( \frac{Q^{2}}{|A^{\tau 4}_{q}|} \right) \right) + \frac{20 C_{A}}{3} \, a_{s}(Q^{2})
\left( \frac{2}{\rho^{2}} \frac{\tilde{I}_{2}(\rho)}{\tilde{I}_{0}(\rho)} + 
\ln \left( \frac{Q^{2}}{|A^{\tau 4}_{q}|} \right) \frac{1}{\rho} \frac{\tilde{I}_{1}(\rho)}{\tilde{I}_{0}(\rho)} \right) 
\right]  \right\},
\end{eqnarray}
\begin{eqnarray}
\frac{1}{e} \, F^{[R]\tau 4,-}_{2}(x,Q^{2}) &=& \frac{32 T_{R} n_{f}}{15 \beta_{0}^{2}} \, f^{\tau 2,-}_{g}(x,Q^{2})
\left\{ A^{\tau 4}_{g} \ln \left( \frac{Q^{2}}{x^{2}_{g}|A^{\tau 4}_{g}|} \right)
-2 C_{A} A^{\tau 4}_{q} \left[ \ln \left( \frac{1}{x_{q}} \right) \ln \left( \frac{Q^{2}}{x_{q}|A^{\tau 4}_{q}|} \right)
- p^{\prime}(\nu_{q}) \right] \right\}\!,
\end{eqnarray}
\end{widetext}
where $a_{s}(Q^{2})\equiv \alpha_{s}(Q^{2})/4\pi$, $x_{a} = x \exp [p(\nu_{a})]$ and $p(\nu_{a}) = \Psi (1 + \nu_{a}) - \Psi (\nu_{a})$. The function 
$\Psi$ is the derivative of the logarithm of the $\Gamma$ function. From quark counting rules we have $\nu_{q} \approx 3$ and $\nu_{g} \approx 4$,
which results in $p(\nu_{q}) \approx 11/6$, $p(\nu_{g}) \approx 25/12$, $p^{\prime}(\nu_{q}) \approx -49/36$, and
$p^{\prime}(\nu_{g}) \approx -205/144$.

In the framework of the infrared renormalon model the twist-six ($\tau 6$) correction to $F_{2}(x,Q^{2})$ is given by \cite{kot2}:
\begin{eqnarray}
F^{[R]\tau 6}_{2}(x,Q^{2}) &=& e \sum_{a=q,g} A^{\tau 6}_{a} \tilde{\mu}^{\tau 6}_{a} (x,Q^{2}) \otimes f^{\tau 2}_{a} (x,Q^{2}) \nonumber \\
 &=& \sum_{a=q,g} F^{[R]\tau 6}_{2,a} (x,Q^{2}) ,
\label{renormalont6}
\end{eqnarray}
where $\tilde{\mu}^{\tau 6}_{a}(x,Q^{2})$ are functions also obtained in \cite{stein001} by means of the infrared renormalon model. Now, taking
into account all higher twist corrections, the structure function $F_{2}$ is given by
\begin{eqnarray}
F^{[R]}_{2}(x,Q^{2}) &=& F^{\tau 2}_{2}(x,Q^{2}) + \frac{1}{Q^{2}}\, F^{[R]\tau 4}_{2}(x,Q^{2}) \nonumber \\
 & & + \frac{1}{Q^{4}}\, F^{[R]\tau 6}_{2}(x,Q^{2}).
\end{eqnarray}

Most importantly, as pointed out in Ref. \cite{kot2}, in the renormalon model the twist-six contribution can be obtained in terms of the
twist-four one:
\begin{eqnarray}
F^{[R]\tau 6}_{2}(x,Q^{2}) = -\frac{8}{7} \left. F^{[R]\tau 4}_{2}(x,Q^{2}) 
\right|_{A^{\tau 4}_{a}\to A^{\tau 6}_{a}, |A^{\tau 4}_{a}|\to \sqrt{|A^{\tau 6}_{a}|}}. \nonumber \\
\label{twistsix08}
\end{eqnarray}

Thus, if $F^{[R]h\tau}_{2}(x,Q^{2})$ denotes the higher-twist operators (twist-4 and twist-6), we have
\begin{eqnarray}
F^{[R]}_{2}(x,Q^{2}) = F^{\tau 2}_{2}(x,Q^{2}) + F^{[R]h\tau}_{2}(x,Q^{2}),
\label{renor001}
\end{eqnarray}
where the ``+'' and the ``-'' representations of $F^{[R]h\tau}_{2}(x,Q^{2})$ can each be put into a compact form:

\begin{widetext}
\begin{eqnarray}
\frac{1}{e} \, F^{[R]h\tau,+}_{2}(x,Q^{2}) &=& \frac{32 T_{R} n_{f}}{15 \beta_{0}^{2}} \, f^{\tau 2,+}_{g}(x,Q^{2})
\sum_{m=4,6} k_{m} \left\{ \frac{A^{\tau m}_{g}}{Q^{(m-2)}} \left( \frac{2}{\rho} \frac{\tilde{I}_{1}(\rho)}{\tilde{I}_{0}(\rho)} + 
\ln \left( \frac{Q^{2}}{|A^{\tau m}_{g}|^{l_{m}}} \right) \right)  \right. \nonumber \\
 & & + \frac{4 C_{F} T_{R} n_{f}}{3 C_{A}} \frac{A^{\tau m}_{q}}{Q^{(m-2)}} \left. \left[ \left(1-\bar{d}^{q}_{+-}(1)a_{s}(Q^{2}) \right) 
\left( \frac{2}{\rho} \frac{\tilde{I}_{1}(\rho)}{\tilde{I}_{0}(\rho)} + 
\ln \left( \frac{Q^{2}}{|A^{\tau m}_{q}|^{l_{m}}} \right) \right) \right. \right.  \nonumber \\
 & & + \left. \left. \frac{20 C_{A}}{3} \, a_{s}(Q^{2}) \left( \frac{2}{\rho^{2}} \frac{\tilde{I}_{2}(\rho)}{\tilde{I}_{0}(\rho)} + 
\ln \left( \frac{Q^{2}}{|A^{\tau m}_{q}|^{l_{m}}} \right) \frac{1}{\rho} \frac{\tilde{I}_{1}(\rho)}{\tilde{I}_{0}(\rho)} \right) 
\right] \right\},
\label{renor002}
\end{eqnarray}
\begin{eqnarray}
\frac{1}{e} \, F^{[R]h\tau,-}_{2}(x,Q^{2}) &=& \frac{32 T_{R} n_{f}}{15 \beta_{0}^{2}} \, f^{\tau 2,-}_{g}(x,Q^{2})
\sum_{m=4,6} k_{m} \left\{ \frac{A^{\tau m}_{g}}{Q^{(m-2)}} \ln \left( \frac{Q^{2}}{x^{2}_{g}|A^{\tau m}_{g}|^{l_{m}}}
\right) \right. \nonumber \\
 & & - \left. 2 C_{A} \frac{A^{\tau m}_{q}}{Q^{(m-2)}} \left[ \ln \left( \frac{1}{x_{q}} \right)
\ln \left( \frac{Q^{2}}{x_{q}|A^{\tau m}_{q}|^{l_{m}}} \right) - p^{\prime}(\nu_{q}) \right] \right\},
\label{renor003}
\end{eqnarray}
\end{widetext}
with $k_{4}=1$, $k_{6}=-8/7$, $l_{4}=1$, and $l_{6}=1/2$. As applied to the case of QCD, from now on we set $N=3$ in order to fix
$C_{A}(=3)$ and
$C_{F}(=4/3)$.

In summary, we have obtained an analytical approach to calculating higher twist corrections to the structure function $F_{2}(x,Q^{2})$.
The formalism is based on existing analytical solutions of the DGLAP equation in the small $x$ region. 
For the present we consider the simplest analytical case, i.e. the generalized DAS approximation with soft initial conditions for parton
distributions. It may be worth noting that analytical approximations to the DGLAP evolution are designed to be valid only for very small
$x$ and/or large $Q^{2}$. However, as we shall see, our analytical approach, when combined with some nonperturbative
information from QCD, results in an instrumental tool to study structure functions also at very slow $Q^{2}$ regime. It is important to
remember that in standard procedures one should solve numerically the DGLAP equation, which makes the procedure cumbersome for
practical use. Thus if we are
only interested in the small $x$ region, there is a clear advantage of the renormalon analytical approach from a practical point of
view.

\section{The QCD effective charge}

It is a commonly accepted view nowadays that the nonperturbative dynamics of QCD may generate an effective momentum-dependent mass
$m(q^{2})$ for the gluons while preserving the $SU(3)_{c}$ local invariance \cite{aguilar001}. Numerical simulations, in which the
space-time continuum is represented as a discrete lattice of points, indicate that such a dynamical mass
does arise when the nonperturbative regime of QCD is probed. Specifically, large-volume lattice QCD simulations, both for
SU(2) \cite{lattice002} and SU(3) \cite{lattice001}, reveal that the gluon propagator is finite in the deep infrared region, both in Landau
gauge and away from it \cite{othergauge001}. In the continuum, it turns out that the nonperturbative dynamics of the gluon propagator is governed
by the corresponding Schwinger-Dyson equations. These equations constitute an infinite set of coupled nonlinear integral
equations governing the dynamics of all QCD Green's functions. According to the Schwinger-Dyson equations, a finite gluon propagator corresponds
to a massive gluon \cite{smekal001}. 

Furthermore, we have known for a long time that the vacuum energy in dynamically broken gauge theories
is related to a dynamical mass \cite{norton001}. Recent studies show that the condition for the
existence of a global minimum of the vacuum energy for a non-Abelian gauge theory, in the presence of a dynamical mass scale, 
implies the existence of a fixed point of the $\beta$ function \cite{aguilar001}. This, in turn, indicates that the QCD exhibits nearly conformal
behavior at infrared momenta \cite{binosi001}.

The QCD effective charge $\bar{\alpha}(q^{2})$ is a nonperturbative generalization of the canonical perturbative running coupling
$\alpha_{s}(q^{2})$ and is intimately related to the phenomenon of dynamical gluon mass generation 
\cite{cornwall001,aguilar002,quinteros001}. The charge $\bar{\alpha}(q^{2})$ provides 
the bridge leading from the deep ultraviolet regime to the deep infrared one. 
It is important to note, however, that the definition of $\bar{\alpha}(q^{2})$ is not unique. 
The effective charge may be obtained, for example, from the ghost-gluon vertex in the Landau gauge \cite{alkofer001}, by considering a
renormalization-group invariant (independent of the renormalization scale $\mu$) quantity $\hat{r}(q^{2})$ defined by
\begin{eqnarray}
\hat{r}(q^{2}) = g^{2}(\mu^{2}) \Delta (q^{2}) F^{2}(q^{2}),
\label{eqan01}
\end{eqnarray}
where $g(\mu^{2})$ is the gauge coupling, $\Delta (q^{2})$ is the gluon propagator and $F^{2}(q^{2})$ is the ghost dressing function. From this
quantity the effective charge may then be defined as 
\begin{eqnarray}
\bar{\alpha}_{gh}(q^{2}) = \left[ q^{2} + m^{2}(q^{2})  \right] \hat{r}(q^{2}),
\label{eqan02}
\end{eqnarray}
where $m(q^{2})$ is the gluon dynamical mass. The relation (\ref{eqan02}) is the definition of the $\bar{\alpha}(q^{2})$ which is most 
commonly used in lattice simulations, since $\Delta (q^{2})$ and $F^{2}(q^{2})$ are quantities measured directly on it.

Alternatively, the QCD effective charge can be obtained within the framework of pinch technique \cite{cornwall001,cornwall002,cornwall003}. This
process-independent definition, obtained from the gluon self-energy $\hat{\Delta}(q^{2})$ in the background-field method \cite{abbott001}, is a
direct non-Abelian generalization of the QED effective charge. In this definition the Schwinger-Dyson solutions for $\hat{\Delta}(q^{2})$ are
used to form another renormalization-group invariant quantity:
\begin{eqnarray}
\hat{d}(q^{2})=g^{2}\hat{\Delta}(q^{2}).
\label{eqan03}
\end{eqnarray}
The inverse of the above quantity may be written
\begin{eqnarray}
\hat{d}^{-1}(q^{2}) = \frac{\left[ q^{2} + m^{2}(q^{2})  \right]}{\bar{\alpha}_{pt}(q^{2})},
\label{eqan04}
\end{eqnarray}
where now
\begin{eqnarray}
\frac{1}{\bar{\alpha}_{pt}(q^{2})} = b_{0} \ln \left( \frac{q^{2} + m^{2}(q^{2})}{\Lambda^{2}}  \right).
\label{eqan05}
\end{eqnarray}
Note that here $b_{0}$ is precisely the first coefficient of the QCD $\beta$ function and $\Lambda$ is the QCD mass scale (of a few hundred MeV).
Despite the distinct theoretical origins of $\bar{\alpha}_{gh}(q^{2})$ and $\bar{\alpha}_{pt}(q^{2})$, they coincide exactly in the deep infrared.
The ultimate reason for this is the existence of a non-perturbative identity relating various of the Green functions appearing in their respective
definitions \cite{quinteros001}.

It is worth remarking on the fact that the form of (\ref{eqan05}) is exactly the same as the form of the leading order (LO) perturbative QCD
coupling, namely
\begin{eqnarray}
\frac{1}{\alpha_{s}^{LO}(p^{2})} = b_{0} \ln \left( \frac{p^{2}}{\Lambda^{2}}  \right),
\label{eqan06}
\end{eqnarray}
if $q^{2} + m^{2}(q^{2}) \to p^{2}$ in the argument of the logarithm; it is this that will effectively ensure that, in practice, the QCD
effective charge can be successfully obtained by saturating the LO perturbative strong coupling $\alpha_{s}^{LO}(q^{2})$. That is to say,
\begin{eqnarray}
\bar{\alpha}^{LO}(q^{2}) &=& \left. \alpha_{s}^{LO}(q^{2}) \right|_{q^{2} \to q^{2} + m^{2}(q^{2})} \nonumber \\
 &=&  \frac{1}{b_{0} \ln \left( \frac{q^{2} + m^{2}(q^{2})}{\Lambda^{2}}  \right)},
\label{eqan07}
\end{eqnarray}
where $b_{0}=\beta_{0}/4\pi = (11C_{A}-2n_{f})/12\pi$. If the Schwinger-Dyson equations preserves the multiplicative renormalizability, 
a next-to-leading order (NLO) effective charge can be built through the same procedure \cite{luna001}:
\begin{eqnarray}
\bar{\alpha}^{NLO}(q^{2}) &=& \frac{1}{b_{0}\ln\left(\frac{q^{2} +
4m^{2}(q^{2})}{\Lambda^{2}}\right)} \nonumber \\
 & & \times \left[1-\frac{b_{1}}{b_{0}^{2}}\frac{\ln\left(\ln\left(\frac{q^{2} +
4m^{2}(q^{2})}{\Lambda^{2}}\right)\right)}{\ln\left(\frac{q^{2} + 4m^{2}(q^{2})}{\Lambda^{2}}\right)} \right] \!\! ,
\label{ansatz2}
\end{eqnarray}
where $b_{1} = \beta_{1}/16\pi^{2} = [34C_{A}^{2} - n_f(10C_{A}+6C_{F})]/48\pi^{2}$. 

We investigate three different types of QCD effective charge $\bar{\alpha}^{NLO}(q^{2})$. They can be constructed from two independent dynamical
gluon masses having distinct asymptotic behaviors: the first runs as an inverse power of a logarithm, the second drops as an
inverse power of momentum \cite{agpapa}. 
A logarithmic running of $m^{2}(q^{2})$ had been initially found in studies of linearized Schwinger-Dyson equations, namely
$m^{2}(q^{2}) \sim \left( \ln q^{2}  \right)^{-1-\gamma}$, with $\gamma > 1$ \cite{cornwall001,aguilar002}. Further studies 
of a non-linear version of the Schwinger-Dyson equation for the gluon self-energy have shown that $m^{2}(q^{2})$ could be rewritten as
\cite{agpapa}
\begin{eqnarray}
m^{2}_{log}(q^{2}) = m_{g}^{2} \left[ \frac{\ln \left( \frac{q^{2}+\rho m_{g}^{2}}{\Lambda^{2}} \right)}{
\ln \left( \frac{\rho m_{g}^{2}}{\Lambda^{2}} \right)} \right]^{-1-\gamma_{1}} ,
\label{eqlog}
\end{eqnarray}
where $\gamma_{1} = -6(1+c_{2}-c_{1})/5$. The parameters $c_{1}$ and $c_{2}$ are related to the ansatz for the three-gluon vertex used in the
numerical analysis of the gluon self-energy. The values of $c_{1}$ and $c_{2}$ are restricted by
a ``mass condition'' that controls the behavior of the dynamical mass in the ultraviolet region, namely $c_{1} \in [0.15,0.4]$ and
$c_{2} \in [-1.07,-0.92]$.  The parameters $m_{g}$ and $\rho$, which control the behavior of $m^{2}_{log}(q^{2})$ in the infrared region, are also
constrained by the mass condition. In particular, they are constrained to lie in the intervals $m_{g} \in [300, 800]$ and
$\rho \in [1.0, 8.0]$ MeV \cite{agpapa}.
A power-law running $m^{2}(q^{2})$ exhibit a distinct behavior whereby, in accordance with OPE
calculations \cite{lav}, the most probable asymptotic behavior of the running gluon mass is proportional to $1/q^2$.
At the level of an non-linear Schwinger-Dyson equation the asymptotic behavior is written as
\begin{eqnarray}
m^{2}_{pl}(q^{2}) = \frac{m_{g}^{4}}{q^{2}+m_{g}^{2}} \left[ \frac{\ln \left( \frac{q^{2}+  \rho m_{g}^{2}}{\Lambda^{2}} \right)}{
\ln \left( \frac{\rho m_{g}^{2}}{\Lambda^{2}} \right)} \right]^{\gamma_{2}-1}  \, ,
\label{eqpo}
\end{eqnarray}
where $\gamma_{2} = (4 + 6 c_{1})/5$. Here the same mass condition imposes $c_{1} \in [0.7,1.3]$ whereas the $\rho$ and
$m_{g}$ parameters are constrained to lie in the same interval as before, namely $\rho \in [1.0, 8.0]$ and
$m_{g} \in [300, 800]$ MeV \cite{agpapa}.

At this point we may start to define the first two QCD effective charges to be explored in this paper. The first charge is constructed simply
by combining the equations (\ref{ansatz2}) and (\ref{eqlog}), henceforth called ``logarithmic charge'' and denoted by
$\bar{\alpha}_{log}(q^{2})$: 
\begin{eqnarray}
\bar{\alpha}_{log}(q^{2}) &=& \frac{1}{b_{0}\ln\left(\frac{q^{2} +
4m^{2}_{log}(q^{2})}{\Lambda^{2}}\right)} \nonumber \\
 & & \times \left[1-\frac{b_{1}}{b_{0}^{2}}\frac{\ln\left(\ln\left(\frac{q^{2} +
4m^{2}_{log}(q^{2})}{\Lambda^{2}}\right)\right)}{\ln\left(\frac{q^{2} + 4m^{2}_{log}(q^{2})}{\Lambda^{2}}\right)} \right] \!\! ,
\label{ansatz3}
\end{eqnarray}

The second one, hereafter called ``power-law charge'' and denoted by $\bar{\alpha}_{pl}(q^{2})$, is
constructed by combining the equations (\ref{ansatz2}) and (\ref{eqpo}):
\begin{eqnarray}
\bar{\alpha}_{pl}(q^{2}) &=& \frac{1}{b_{0}\ln\left(\frac{q^{2} +
4m^{2}_{pl}(q^{2})}{\Lambda^{2}}\right)} \nonumber \\
 & & \times \left[1-\frac{b_{1}}{b_{0}^{2}}\frac{\ln\left(\ln\left(\frac{q^{2} +
4m^{2}_{pl}(q^{2})}{\Lambda^{2}}\right)\right)}{\ln\left(\frac{q^{2} + 4m^{2}_{pl}(q^{2})}{\Lambda^{2}}\right)} \right] \!\! .
\label{ansatz3}
\end{eqnarray}

The third effective charge can be postulated by considering renormalization aspects of Lagrangians containing an explicit mass term for the
vector field.
As demonstrated by Curci and Ferrari some time ago, the action obtained from these massive Yang-Mills Lagrangians, while also containing a
four-vertex for the Faddeev-Popov ghost field, is invariant under a generalized nonlinear gauge transformation \cite{Curci76}. 
The invariance under this transformation, which is an extended version of the BRST one, guarantees the validity of the Slavnov-Taylor
identities. More recently, the two-point correlation functions of gluons and ghosts for the pure Yang-Mills theory in Landau gauge were
accurately reproduced for all momenta by using a particular case of the Curci-Ferrari Lagrangians \cite{Tissier:2010ts,Tissier:2011ey}, i.e.
the one-loop calculation successfully reproduces lattice simulations in $d=4$.
More important, the introduction of a bare gluon mass in the gauge-fixed
Lagrangian allows to capture the main effects of Gribov copies in a reliable way \cite{Serreau:2012cg,Tissier:2017fqf}.
Furthermore, two- and three-point correlation functions have been computed within this massive model and also successfully compared with lattice
simulations \cite{Pelaez:2013cpa}. The introduction of two-loop corrections improve the comparison of ghost and gluon
two-point correlation functions with lattice data in the quenched approximation \cite{Gracey:2019xom}. 
The Curci-Ferrari Lagrangian is infrared safe, meaning that there exists renormalization schemes without Landau pole. 
In particular, the one-loop calculation implies that the $\beta$ function for the coupling constant $g$ in the infrared region behaves like
\begin{eqnarray}
\beta_{g} \sim c_{0} \, \frac{g^{3}}{4\pi},
\end{eqnarray}
where $c_0=C_A/24\pi$. This means that the strong coupling vanishes logarithmically in the infrared, in agreement with some recent lattice
results using a renormalization group invariant coupling resulting from a particular combination of the gluon and ghost propagators
\cite{Duarte:2016iko,Oliveira:2016stx,Zafeiropoulos:2019flq}. Thus, our Curci-Ferrari effective charge can be constructed as follows:
\begin{eqnarray}
\bar{\alpha}_{_{CF}}(q^{2}) = \frac{1}{1+c_{0}\ln\left(1 + \frac{4m^{2}_{log}(q^{2})}{q^{2}}\right)} \, \bar{\alpha}_{log}(q^{2}).
\label{ansatzCF}
\end{eqnarray}

It may be worth emphasizing that the QCD effective charges $\bar{\alpha}_{log}(q^{2})$, $\bar{\alpha}_{pl}(q^{2})$ and $\bar{\alpha}_{_{CF}}(q^{2})$
exhibit infrared fixed points as $q^2\rightarrow  0$, i.e. the dynamical gluon mass tames the Landau pole. Moreover, in the limit
$q^2 \gg \Lambda^2$ these effective charges match with the canonical perturbative two-loop coupling:
$\bar{\alpha}(q^2 \gg \Lambda^2)\to \alpha_{s}(q^{2})$.
The analyticity of $\bar{\alpha}_{log}(q^{2})$, $\bar{\alpha}_{pl}(q^{2})$ and $\bar{\alpha}_{_{CF}}(q^{2})$
is automatically preserved if the gluon mass scale is set larger than half of the QCD scale parameter, namely $m_{g}/\Lambda >1/2$ \cite{luna002}.
This ratio may be phenomenologically determined \cite{luna001,luna002,luna003} and typically lies in the interval $m_{g}/\Lambda \in [1.1, 2]$.
The canonical coupling $\alpha_{s}(q^{2})$, on the other hand, has Landau singularities on the spacelike semiaxis
$0 \leq q^{2} \leq \Lambda ^2$. That is, $\alpha_{s}$ has a nonholomorphic behavior at low $q^{2}$ \cite{stefanis001}.
This problem has been worked out by using analytic versions of QCD whose coupling $\alpha_{s}(q^{2})$ is holomorphic in the entire complex
plane except the timelike axis ($q^{2}<0$) \cite{cvetic007}, with many applications in hadronic physics \cite{cvetic004,cvetic005}. In
a mathematical sense, the QCD effective charges belong to the same class of holomorphic couplings.
Moreover, as pointed out by Cveti\v{c} \cite{cvetic004}, evaluation of renormalization scale-invariant
quantities at low $Q^{2}$, in terms of infrared finite couplings, can be effectively done as a series in
derivatives of the coupling with respect to the logarithm of $Q^{2}$. This truncated series exhibit significantly better convergence properties.

\section{Results}

The nucleon structure function $F_{2}(x,Q^{2})$
has been measured in deep inelastic scattering (DIS) of leptons off nucleons at the HERA collider.
In this work we carry out global fits to small-$x$ $F_{2}(x,Q^{2})$ data at low and moderate $Q^{2}$ values \cite{heradata}, where $x$ is the
Bjorken variable and $Q^{2}$ is the virtuality of the photon.
We use HERA data from the ZEUS and H1 Collaborations, with the statistic and systematic errors added in quadrature.
Specifically, we fit to the structure function at $Q^{2}=0.2$, 0.25, 0.3, 0.5, 0.65, 0.85, 1.2, 1.3, 1.5, 1.9, 2.0, 2.5, 3.5, 5.0, 6.5 and 10
GeV$^{2}$. 
The global fits were performed using a $\chi^{2}$ fitting procedure, where the value of $\chi^{2}_{min}$ is distributed as a $\chi^{2}$ distribution
with $\nu$ degrees of freedom. We have adopted an interval $\chi^{2}-\chi^{2}_{min}$ corresponding to the projection
of the $\chi^{2}$ hypersurface enclosing 90\% of probability. As test of goodness-of-fit we adopt the chi-square per degree of freedom,
namely $\tilde{\chi} \equiv \chi^2/\nu$.

In our global analysis we have chosen to use only $F_{2}$ structure function data. With regard to this choice, it is known that for
inclusive $e^{\pm}p$ DIS process the real experimentally measured data are the reduced cross sections
$\tilde{\sigma}$,
\begin{eqnarray}
\frac{d^{2}\sigma^{e^{\pm}p} }{dx\, dQ^{2}} = \frac{2\pi \alpha^{2} Y_{+}}{xQ^{4}} \, \tilde{\sigma}(x,Q^{2},y),
\label{reduced001}
\end{eqnarray}
where $y$ is the inelasticity, $\alpha$ is the fine structure constant and $Y_{+} = 1 + (1-y)^{2}$. The reduced cross section, at low virtuality
of the exchanged boson, is expressed in terms of 2 structure functions, namely $F_{2}$ and $F_{L}$:
\begin{eqnarray}
\tilde{\sigma}(x,Q^{2},y) = F_{2}(x,Q^{2})   -\frac{y^{2}}{Y_{+}} F_{L}(x,Q^{2}). 
\label{reduced002}
\end{eqnarray}
The kinematic variables in (\ref{reduced001})-(\ref{reduced002}) are related via $Q^{2} = xys$, where $\sqrt{s}$ is the $ep$ center-of-mass (CM)
energy. The longitudinal structure function $F_{L}$
is proportional to the longitudinally polarized virtual-photon absorption cross section, $F_{L} \propto \sigma_{L}$, whereas $F_{2}$ includes also
the the absorption cross section for transversely polarized virtual-photons, $F_{2} \propto (\sigma_{L} + \sigma_{T})$. The ratio
$R = \sigma_{L}/\sigma_{T} = F_{L}/(F_{2}-F_{L})$ gives the relative strength of the two structure functions. Our analyses is based on a sample of
HERA data restricted to the kinematic region where the contribution of $F_{L}$ on the reduced cross section is relatively small \cite{heradata}.
For this
selected data set, which has been measured at essentially a fixed CM energy, $F_{L}$ is usually a calculated correction. For example, in one of
the ZEUS analyses, to correct for the effect of $F_{L}$ on $\tilde{\sigma}$, the value of $R$ from the BKS model \cite{bks001} was used, which
was parameterized, to a good approximation, by $R = 0.165 \, Q^{2}/m^{2}_{\rho}$, where $m_{\rho} = 0.77$ GeV. This correction changed the extracted
$F_{2}$ values by at most 3\% with respect to the values determined putting $F_{L} = 0$. In another study, $F_{2}$ was determined by ZEUS assuming
$\sigma_{L}$ to be the value given by the vector dominance model, $\sigma_{L} = K (Q^{2}/m^{2}_{\rho}) \sigma_{T}$, where $K = 0.5$. The effect on
$F_{2}$ was typically around 1-2\% for most $x$ values, and increased $F_{2}$ by up to 7\% in the lowest $x$ values. In the case of H1 analyses,
in order to extract $F_{2}$ from the reduced cross section, $R$ values calculated according to the QCD prescription \cite{martinelli887} (using
the MRS PDF \cite{mrsd001}) were chosen. These values reduced the cross section by at most 8\% with respect to the values determined assuming
$F_{L} = 0$. Hence, in our analysis, the bias introduced by neglecting $F_{L}$ is kept to a minimum.

A suitable combination of the asymptotic freedom and the factorization properties of QCD results in a systematic expansion of the DIS cross
sections in terms of $\alpha_{s}(Q^{2})$, evaluated at the virtuality scale. Thus the virtuality $Q$ is the natural scale in our calculations.
This means that all the couplings $\alpha_{s}(Q^{2})$ appearing in the Section II are replaced by QCD effective charges evaluated at the virtuality
scale: $\alpha_{s}(Q^{2}) \to \bar{\alpha}_{log}(Q^{2})$, $\bar{\alpha}_{pl}(Q^{2})$ or $\bar{\alpha}_{_{CF}}(Q^{2})$ (or, equivalently, that $q^{2}$
is replaced by $Q^{2}$ in Eqs. (50)-(55)). 
In all the fits we fix $n_{f}=3$ and $\Lambda = 284$ MeV. The latter is not only consistent to NLO procedures, but is also the same one adopted
in Refs. \cite{luna001} and \cite{cve}. The former choice is justified by the fact that most of the data lie at $Q$ values below the charm mass
$m_{c}$. In the $n_{f}=3$ scheme the charm can only be pair-produced by gluon splittings when kinematically allowed.
At the low scales of the data considered in the fits, production of charm is not kinematically allowed and therefore in the $n_{f}=3$ scheme the
charm quark decouples.

Concerning the dynamical masses (\ref{eqlog}) and (\ref{eqpo}), in both cases we set $\rho=4$; it is the optimal value originally
obtained by Cornwall in order to reproduce numerical results of a Schwinger-Dyson equation for the gluon propagator
\cite{cornwall001}. Moreover, this $\rho$ value is consistent with the expected threshold $Q^{2}=4m^{2}_{g}$ for gluons to pop up from the vacuum
\cite{cornwall004}. As in the previous study \cite{luna001}, we set $\gamma_{1} = 0.084$ and $\gamma_{2} = 2.36$ since again the best
determinations
of $F_{2}$ come from the analyses using these values. We have observed this general pattern in all analysis performed in this work, meaning that
the magnitudes of $\gamma_{1}$ and $\gamma_{2}$ are not sensitive to twist corrections and changes in the form of the QCD effective charges.

\begin{table}
\centering
\caption{The values of the fitting parameters from the global fit to $F_{2}$ data. Results obtained using the logarithmic effective charge.}
\begin{tabular}{cccc}
  & $\tau 2$ & $\tau 2 + \tau 4$ & $\tau 2 + \tau 4 + \tau 6$  \\
\hline \\[-2.2ex]
$m_{g}$ [MeV]   & 340$\pm$17 & 284$\pm$17  &  310$\pm$53 \\
$Q_{0}^{2}$ [GeV$^{2}$] & 0.080$\pm$0.048 & 0.54$\pm$0.17  &  0.99$\pm$0.16  \\
$A_{g}$ & 0.091$\pm$0.070 & 0.42$\pm$0.24  &  1.19$\pm$0.26  \\
$A_{q}$ & 0.727$\pm$0.054 &  0.60$\pm$0.12  &  0.422$\pm$0.086  \\
$A^{\tau 4}_{g}$ & - & 0.59$\pm$0.26  &  0.58$\pm$0.19  \\
$A^{\tau 4}_{q}$ & - & 0.020$\pm$0.018  &  0.232$\pm$0.081  \\
$A^{\tau 6}_{g}$ & - & -  &  0.139$\pm$0.076 \\
$A^{\tau 6}_{q}$ & - & -  &  0.0203$\pm$0.0082 \\[0.5ex]
\hline \\[-2.2ex]
$\nu$ & 246  & 244  & 242 \\
$\tilde{\chi}$ & 2.41  & 2.08  & 1.21 \\
\end{tabular}
\end{table}
\begin{table}
\centering
\caption{The values of the fitting parameters from the global fit to $F_{2}$ data. Results obtained using the power-law effective charge.}
\begin{tabular}{cccc}
  & $\tau 2$ & $\tau 2 + \tau 4$ & $\tau 2 + \tau 4 + \tau 6$  \\
\hline \\[-2.2ex]
$m_{g}$ [MeV]   & 360$\pm$9 & 282$\pm$24  &  415$\pm$67 \\
$Q_{0}^{2}$ [GeV$^{2}$] & 0.11$\pm$0.15 & 0.929$\pm$0.073  &  1.17$\pm$0.19  \\
$A_{g}$ & -0.090$\pm$0.031 & 0.856$\pm$0.080  &  1.37$\pm$0.34  \\
$A_{q}$ & 0.857$\pm$0.017 &  0.488$\pm$0.042  &  0.403$\pm$0.081  \\
$A^{\tau 4}_{g}$ & - & 0.69$\pm$0.14  &  0.39$\pm$0.30  \\
$A^{\tau 4}_{q}$ & - & 0.132$\pm$0.013  &  0.38$\pm$0.16  \\
$A^{\tau 6}_{g}$ & - & -  &  0.135$\pm$0.073 \\
$A^{\tau 6}_{q}$ & - & -  &  0.040$\pm$0.014 \\[0.5ex]
\hline \\[-2.2ex]
$\nu$ & 246  & 244  & 242 \\
$\tilde{\chi}$ & 2.88  & 1.38  & 1.19 \\
\end{tabular}
\end{table}

We start our investigation by considering the twist-two (leading-twist) expansion of $F_{2}(x, Q^2)$. We determine the values of the
parameters $m_{g}$, $Q_0^2$, $A_g$ and $A_q$ using the effective charges $\bar{\alpha}_{log}(Q^{2})$, $\bar{\alpha}_{pl}(Q^{2})$ and
$\bar{\alpha}_{_{CF}}(Q^{2})$. The values of the fitted parameters are given in Tables I, II and III, in the columns headed ``$\tau 2$''. 
The values of  $\tilde{\chi}$ in the case of logarithmic, power-law and Curci-Ferrari charges are 2.41, 2.88 and 2.39, respectively, and the
structure functions corresponding to these values are
shown by the dotted curves in Figures 1, 2 and 3. It is clear from the relatively high values of $\tilde{\chi}$ obtained in these global fits,
as well as from the dotted curves depicted in the Figures, that the leading twist ($\tau 2$) expansion provides poor fits for the $x$
dependence of $F_{2} (x, Q^2)$ for specific $Q^2$ bins, especially in the deep infrared region (data in the region 0.2
GeV$^{2}$ $\lesssim Q^{2}  \lesssim 0.5$ GeV$^{2}$).

Some improvement in the $\tilde{\chi}$ values is achieved by incorporating the twist-four ($\tau 4$) correction to $F_{2}$. 
In this case the $\tilde{\chi}$ values obtained using logarithmic, power-law and Curci-Ferrari charges are 2.08, 1.38 and 1.34, respectively. 
In the combination of the leading and the twist-four correction we have 6 free parameters: $m_{g}$, $Q_0^2$, $A_g$, $A_q$, $A^{\tau 4}_{g}$ and
$A^{\tau 4}_{q}$. Their values are given in Tables I, II and III, in the columns headed ``$\tau 2 + \tau 4$''.
However, as shown by the dashed curves in Figures 1, 2 and 3, the fits are not so good, especially in the deep infrared region.
Thus the improvement in the description of $F_{2}$ is not enough, mostly because the flat trend of the infrared data.

\begin{table}
\centering
\caption{The values of the fitting parameters from the global fit to $F_{2}$ data. Results obtained using the Curci-Ferrari effective charge.}
\begin{tabular}{cccc}
  & $\tau 2$ & $\tau 2 + \tau 4$ & $\tau 2 + \tau 4 + \tau 6$  \\
\hline \\[-2.2ex]
$m_{g}$ [MeV]   & 326$\pm$72 & 234$\pm$14 &  302$\pm$53 \\
$Q_{0}^{2}$ [GeV$^{2}$] & 0.05$\pm$1.35 & 0.883$\pm$0.071  &  0.97$\pm$0.16  \\
$A_{g}$ & 0.09$\pm$0.30 & 0.846$\pm$0.075  &  1.19$\pm$0.28  \\
$A_{q}$ & 0.73$\pm$0.31 &  0.491$\pm$0.041  &  0.420$\pm$0.091  \\
$A^{\tau 4}_{g}$ & - & 0.65$\pm$0.13  &  0.55$\pm$0.19  \\
$A^{\tau 4}_{q}$ & - & 0.1179$\pm$0.0090  &  0.224$\pm$0.082  \\
$A^{\tau 6}_{g}$ & - & -  &  0.131$\pm$0.076 \\
$A^{\tau 6}_{q}$ & - & -  &  0.0194$\pm$0.0078 \\[0.5ex]
\hline \\[-2.2ex]
$\nu$ & 246  & 244  & 242 \\
$\tilde{\chi}$ & 2.39  & 1.34  & 1.20 \\
\end{tabular}
\end{table}

Clearly, we can improve on the description of the structure function in the deep infrared region by adding the twist-six ($\tau 6$) correction to
the previous case.
Now, the $\tilde{\chi}$ values obtained using logarithmic, power-law and Curci-Ferrari charges are 1.21, 1.19 and 1.20, respectively.
We have in this analysis 8 fitting parameters: $m_{g}$, $Q_0^2$, $A_g$, $A_q$, $A^{\tau 4}_{g}$, $A^{\tau 4}_{q}$, $A^{\tau 6}_{g}$ and $A^{\tau 6}_{q}$.
Their values are given in Tables I, II and III, in the columns headed ``$\tau 2 + \tau 4 + \tau 6$''. The low values obtained for $\tilde{\chi}$
indicate good agreement between the theory and the experimental data, as indeed shown by the solid curves in Figures 1, 2 and 3. 
In Figure 4 we compare the higher-twist results in the deep infrared region.

In particular, for the mass scales obtained by using $\bar{\alpha}_{log}(Q^{2})$, $\bar{\alpha}_{pl}(Q^{2})$ and $\bar{\alpha}_{_{CF}}(Q^{2})$
effective charges, we have $m_{g}=310\pm 53$ MeV, $m_{g}=415\pm 67$ MeV and $m_{g}=302\pm 53$ MeV, respectively. These mass values preserve the
analyticity constraint on the strong running coupling, since for the three values we have $m_{g}/\Lambda > 1/2$. The resulting QCD effective
charges are shown in Figure 5. The plot shows that  since the effective coupling is less than
one, then the use of fixed-order perturbation theory is (almost) justified, however since the coupling is large, 
higher order corrections beyond NLO are expected to be large; even so, the use of a finite QCD effective charge as the expansion parameter
provides a basis for regulating the infrared nonperturbative domain of $\alpha_{s}$. Specifically, there is the possibility to
build up a skeleton expansion where the nonperturbative information would be included in propagators and vertices \cite{brodskychineses}. In this
way the finite nature of the QCD running coupling constant
at low energy scales could allow to capture at an inclusive level the nonperturbative effects
in a reliable way.

We can now calculate the gluon distribution function $f_{g}(x,Q^{2})=x g(x,Q^{2})$, since in the small-$x$ regime the gluon is by far the
dominant contribution to $F_{2}$. Note that in the expression (\ref{f211}) the parton distributions $f_{q}$ and $f_{g}$ contain both high twist
corrections, so that
\begin{eqnarray}
f_{a}(x,Q^{2}) &=& f^{\tau 2}_{a}(x,Q^{2}) + \frac{1}{Q^{2}}\, f^{\tau 4}_{a}(x,Q^{2}) \nonumber \\
 & & + \frac{1}{Q^{4}}\, f^{\tau 6}_{a}(x,Q^{2}),
\label{56a} 
\end{eqnarray}
where $a=q, g$. The expressions (\ref{f211}), (\ref{renormalont4}), (\ref{renormalont6}) and (\ref{56a}) are consistent with the equation
\begin{eqnarray}
f^{\tau 4,\tau 6}_{g}(x,Q^{2}) &=& \frac{c_{1}}{a_{s}(Q^{2})} \, A^{\tau 4,\tau 6}_{g} \tilde{\mu}^{\tau 4,\tau 6}_{g} (x,Q^{2}) \otimes
f^{\tau 2}_{g} (x,Q^{2}) \nonumber \\
 &=& \frac{1}{e} \frac{c_{1}}{a_{s}(Q^{2})} \, F^{[R]\tau 4,\tau 6}_{2,g} (x,Q^{2}) ,
\label{renormalong1}
\end{eqnarray}
where $c_{1}=3/4 T_{R}n_{f}$, and hence the twist-four correction $f^{\tau 4}_{g}(x,Q^{2})$ to the gluon distribution function can be written as
\begin{eqnarray}
f^{\tau 4}_{g}(x,Q^{2}) = f^{\tau 4,+}_{g}(x,Q^{2}) + f^{\tau 4,-}_{g}(x,Q^{2}),
\end{eqnarray}
\begin{widetext}
\begin{eqnarray}
\frac{f^{\tau 4,+}_{g}(x,Q^{2})}{f^{\tau 2,+}_{g}(x,Q^{2})} = \frac{8}{5 \beta_{0}^{2}} \frac{A^{\tau 4}_{g}}{a_{s}(Q^{2})} \,
\left[ \frac{1}{Q^{2}} \left( \frac{2}{\rho} \frac{\tilde{I}_{1}(\rho)}{\tilde{I}_{0}(\rho)}
+ \ln \left( \frac{Q^{2}}{|A^{\tau 4}_{g}|} \right) \right) \right] + {\cal O}(\rho),
\end{eqnarray}
\begin{eqnarray}
\frac{f^{\tau 4,-}_{g}(x,Q^{2})}{f^{\tau 2,-}_{g}(x,Q^{2})} = \frac{8}{5 \beta_{0}^{2}} \frac{A^{\tau 4}_{g}}{a_{s}(Q^{2})} \, 
\left[ \frac{1}{Q^{2}} \, \ln \left( \frac{Q^{2}}{x^{2}_{g}|A^{\tau 4}_{g}|} \right) \right] + {\cal O}(x),
\end{eqnarray}
\end{widetext}
whereas its twist-six correction $f^{\tau 6}_{g}(x,Q^{2})$ can be obtained from (\ref{twistsix08}) merely changing $F_{2}^{[R]\tau 4}$ and
$F_{2}^{[R]\tau 6}$ to $f_{g}^{\tau 4}$ and $f_{g}^{\tau 6}$, respectively:
\begin{eqnarray}
f^{\tau 6}_{g}(x,Q^{2}) = -\frac{8}{7} \left. f^{\tau 4}_{g}(x,Q^{2}) 
\right|_{A^{\tau 4}_{g}\to A^{\tau 6}_{g}, |A^{\tau 4}_{g}|\to \sqrt{|A^{\tau 6}_{g}|}}. \nonumber \\
\end{eqnarray}

In Figure 6 we compare our gluon distribution functions (GDFs), calculated using $\bar{\alpha}_{pl}(Q^{2})$ and
$\bar{\alpha}_{CF}(Q^{2})$, with
two new generation PDFs, namely CT14 \cite{ct14001} and MMHT \cite{mmht001} sets. These PDFs include updated data from Tevatron and HERA
experiments as well as data from the LHC, such as inclusive production of jets \cite{ct14jet} and vector bosons
\cite{ct14boson} at $\sqrt{s} = 2.76$, 7 and 8 TeV. 

Comparing the renormalon and standard
GDFs, we see that our distributions $f_{g}(x, Q^{2})$ are in good agreement with the CT14 and MMHT ones at very small $x$ (the results are within
the uncertainty band of the MMHT gluon distribution). As expected, the smaller the value of $x$, the better the agreement. The present level of
agreement follows directly after noting that the very small $x$ region is just the region where the generalized double-asymptotic-scaling
approximation is fully satisfied. On the other hand, the agreement tend to get worse for higher $x$ values: namely, at $x\sim 10^{-2}$, while at
low $Q^{2}$ values the differences in the individual distributions are typically of the order of a few per cent, at $Q^{2}=10^{3}$ GeV$^{2}$ the
difference between the renormalon and the standard GDFs is already a factor of 2. Clearly our GDFs will overshoot the corresponding CT14 and MMHT
distributions in the medium-large $x$ region. Since our analysis uses the $n_{f}=3$ scheme, our results have been compared with CT14 and MMHT
gluon distributions obtained using fixed flavour number scheme, namely CT14nlo$\_$NF3 and mmht2014nlo118$\_$nf3 sets, respectively.
We have calculated the uncertainty in the gluon distribution in the case of the MMHT set. The uncertainty band has been obtained from the errors
analyses as described in the respective MMHT papers by summing over the PDFs given in the eigenvector sets. 

The reason for this discrepancy becomes clear
when we consider how standard PDFs are obtained. They are determined from global analyses of the available deep inelastic and hard
scattering data, using parameterizations for the input distributions (at an initial scale $Q_{0}^{2}$) sufficiently smooth and flexible enough to
accommodate all of the experimental data included in the global fits. The parameterization is evolved to any other scale
$Q^{2} \ge Q_{0}^{2}$ via DGLAP equations \cite{dglap}. In practice, most groups adopt a functional form
\begin{eqnarray}
xf_{i}(x, Q_{0}^{2}) = A_{i}\, x^{\delta_{i}} (1 - x)^{\eta_{i}} F (x, \{ \gamma_{i} \}) ,
\label{prmtrs001}
\end{eqnarray}
where $i$ is the flavor label (including the gluon) and $F (x, \{ \gamma_{i} \})$ is a smooth function which remains finite in the limits
$x\to 0$ and $x\to 1$. The parameters $A_{i}$, $\delta_{i}$, $\eta_{i}$ and the set
of factors $\{ \gamma_{i} \}$ are then determined from the global fits. The requirement that $xf_{i}(x, Q^{2})$ vanishes in the elastic limit
$x \to 1$ turns out to be of central importance in the large $x$ region. This constraint has its origin in the Brodsky-Farrar counting
rule \cite{bf001} 
\begin{eqnarray}
\left. xf_{i}(x, Q^{2}) \right|_{x \to 1}  = (1 - x)^{\eta_{i}},
\end{eqnarray}
already included in the expression (\ref{prmtrs001}) and in broad agreement, within the PDFs uncertainties, with the global fit results.
At this stage we have to remember that the analytical formalism we have been discussing in the context of the generalized
double-asymptotic-scaling is formulated in terms of analytical solutions of the DGLAP equation valid only in the small-$x$ limit, leading to
parton distribution functions having soft initial conditions; it is therefore not possible to introduce any kinematic constraint at all beyond the
small $x$ region. This immediately raises an important question: is it possible to relax the soft condition (\ref{eq1}) by adopting a more flexible
parton initial condition and repeat the analysis including $F_{2}$ data at medium-large $x$ region? The answer is yes, and that will be the next
step of our program. However, as discussed in Section 2, in this paper we are only interested in the small $x$ region;
a further central aim of this paper
is to test the applicability of the simplest analytical version of the DGLAP solution, when combined with nonperturbative information from the QCD
in order to extend the approach to the infrared region (low $Q^{2}$ region). A cumbersome numerical analysis covering
a larger range in $x$ is therefore well beyond the scope of this paper.

\section{Conclusions}

In this letter we have investigated the structure function $F_{2}(x,Q^{2})$ of the proton by means of the {\it generalized} DAS
approximation. We have studied the effect of higher twist operators of the Wilson operator product expansion in $F_{2}(x,Q^{2})$ at small-$x$.
The higher twist corrections have been obtained from the infrared renormalon model in which the higher twist operators can be expressed in
terms of the leading-twist $F_{2}$ expansion.

In order to describe the infrared $F_{2}$ data we have introduced three different QCD effective charges at NLO. The charges have been
constructed by using a phenomenological procedure that mimics the process-independent method of obtaining effective charges within the
framework of pinch technique. One of the QCD charges, denoted $\bar{\alpha}_{_{CF}}(Q^{2})$, is postulated by considering renormalization aspects of
massive Yang-Mills Lagrangians and it goes to zero in the limit $Q^{2} \to 0$. The other two, namely $\bar{\alpha}_{log}(q^{2})$ and
$\bar{\alpha}_{pl}(q^{2})$, have distinct infrared behaviors but similar
freezing processes when the $Q^{2}$ scale goes to 0. The three charges exhibit infrared fixed points as $Q^2\rightarrow  0$, i.e. the dynamical
gluon mass $m(Q^{2})$ tames the Landau pole. Moreover, these charges are examples of holomorphic running couplings, since they preserve the
analyticity constraint in any arbitrary scale $Q$.

We have performed global fits to $F_{2}$ data using either its leading twist expansion or its higher-twist version. We have observed that the
best fits are obtained when we introduce the higher-twist corrections to $F_{2}$. In fact there is a large improvement in the description of the
structure function in the deep infrared region only when the $\tau 4$ and $\tau 6$ corrections are introduced at the same time. In other words,
the fitting parameters are very strongly constrained by the deep infrared data sets. As indicated in the previous section, when all higher-twist
corrections are taken into account, the gluon masses are found to be equal to $m_{g}=310$ MeV, $m_{g}=415$ MeV and $m_{g}=302$ MeV for the case of
logarithmic, power-law and Curci-Ferrari effective charges, respectively. It is noteworthy that these $m_{g}$ values are of the same order of
magnitude as the gluon masses obtained in other calculations of strongly interacting processes \cite{halzen,luna001,luna002}.

Furthermore, a nearly two-decade long effort to compute small-$x$ resummation \cite{ball021}, taking into account simultaneous resummation of
collinear as well as anti-collinear singularities in the evolution kernels, has recently culminated in the resummation of small-$x$ logarithms
in the PDFs via a full consistent use of BFKL in conjunction with DGLAP \cite{ball022}. This novel method of small-$x$ resummation stabilizes
the perturbative expansion of DIS structure functions at small values of $x$ and $Q^{2}$, leading to a much less step behavior of the
unintegrated gluon distribution at low $Q^{2}$. According to this picture, a steeply-rising gluon distribution may be absent at very low $Q^{2}$
scales and, as $Q^{2}$ increases, be generated radiatively through perturbative evolution. It raises the possibility that it may be possible the
existence of a flat input distribution at infrared momenta. We argue that all these results corroborate theoretical analyzes
considering the nonperturbative phenomenon of dynamical mass generation in QCD.

Finally, although $F_{L}$ is usually treated as a small correction in the $F_{2}$ extraction from the reduced cross section, it is an important
quantity due to its rather direct relation to $f_{g}(x,Q^{2})$.
Thus, it is clearly important to develop a consistent QCD method to describe directly the full cross section $\tilde{\sigma}$. Work in this
direction, using the renormalon approach, is in progress.

\section*{Acknowledgments}
We are grateful to N. Wschebor and G. Hern\'andez-Chifflet for valuable discussions.
This research was partially supported by the PEDECIBA program and by the ANII-FCE-1-126412 project.

\newpage

\begin{figure*}
\vspace{1.0cm}
\begin{center}
\includegraphics[height=.70\textheight]{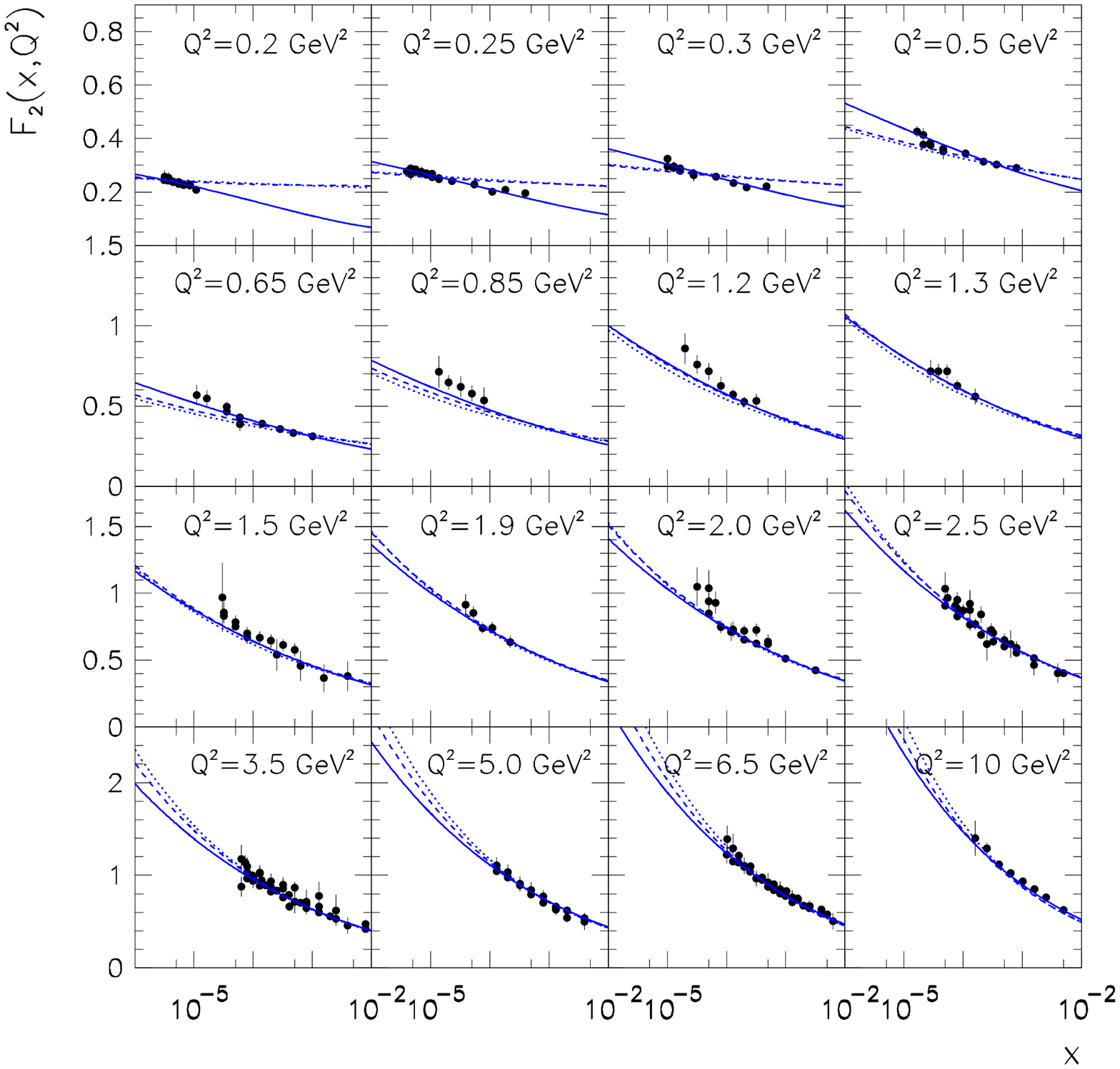}
\caption{Fits of the $x$ dependence of $F_2 (x, Q^2)$ for specific $Q^2$ with the {\it logarithmic} effective charge. The dotted curves
correspond to the twist-two (leading) approximation. The dashed (solid) curves were obtained considering the twist-four (twist-six)
correction to $F_2 (x, Q^2)$.}
\label{fig1}
\end{center}
\end{figure*}

\begin{figure*}
\vspace{1.0cm}
\begin{center}
\includegraphics[height=.70\textheight]{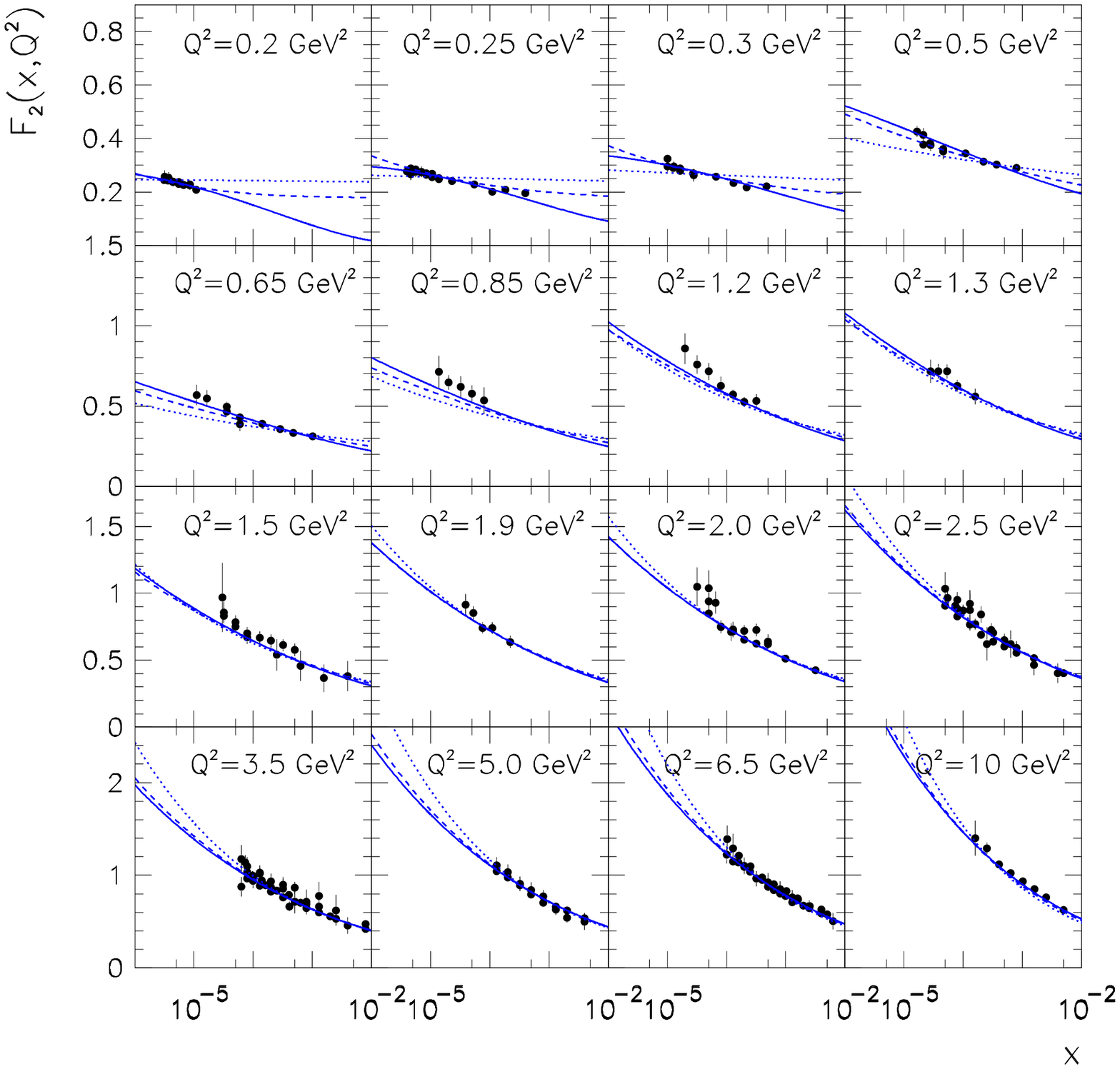}
\caption{Fits of the $x$ dependence of $F_2 (x, Q^2)$ for specific $Q^2$ with the {\it power-law} effective charge. The dotted curves
correspond to the twist-two (leading) approximation. The dashed (solid) curves were obtained considering the twist-four (twist-six)
correction to $F_2 (x, Q^2)$.}
\label{fig1}
\end{center}
\end{figure*}

\begin{figure*}
\vspace{1.0cm}
\begin{center}
\includegraphics[height=.70\textheight]{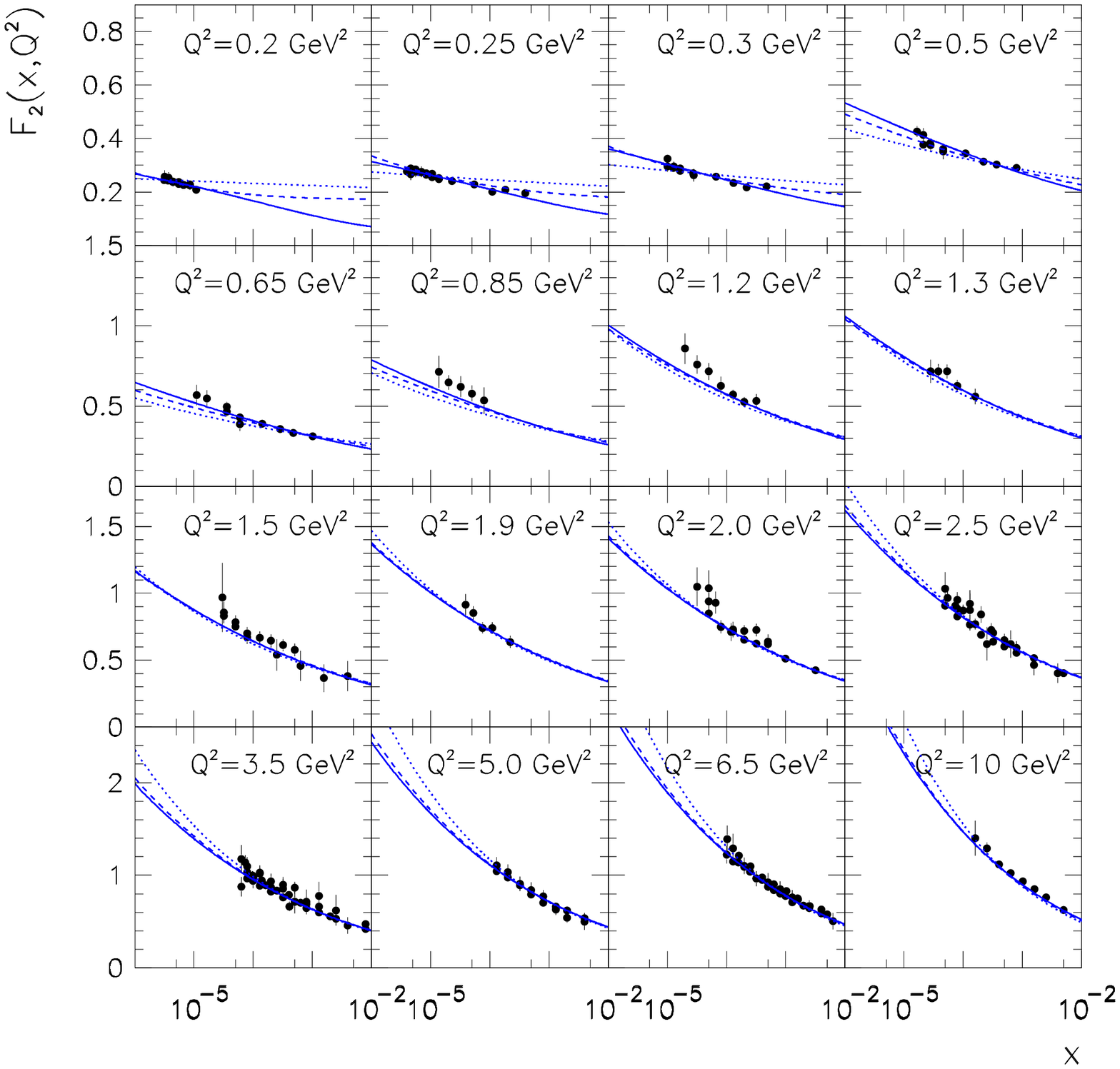}
\caption{Fits of the $x$ dependence of $F_2 (x, Q^2)$ for specific $Q^2$ with the {\it Curci-Ferrari} effective charge. The dotted curves
correspond to the twist-two (leading) approximation. The dashed (solid) curves were obtained considering the twist-four (twist-six)
correction to $F_2 (x, Q^2)$.}
\label{fig1}
\end{center}
\end{figure*}

\begin{figure*}
\vspace{1.0cm}
\begin{center}
\includegraphics[height=.70\textheight]{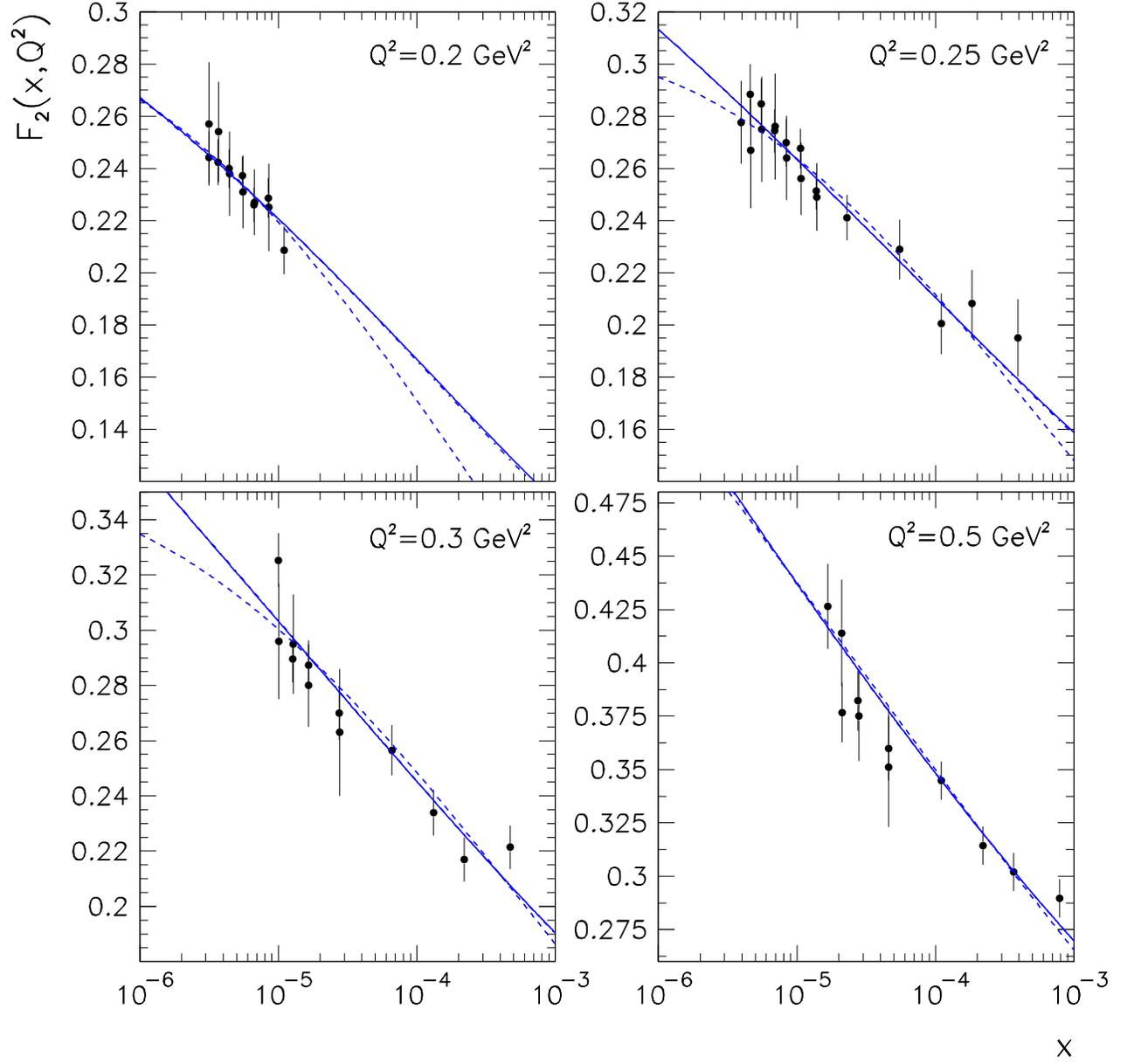}
\caption{Fits of the $x$ dependence of $F_2 (x, Q^2)$ for specific $Q^2$ in the deep infrared region. 
The dotted-dashed, the solid and the dashed curves correspond to the logarithmic, the Curci-Ferrari and the power-law effective charges, respectively.}
\label{fig2}
\end{center}
\end{figure*}

\begin{figure*}
\vspace{1.0cm}
\begin{center}
\includegraphics[height=.70\textheight]{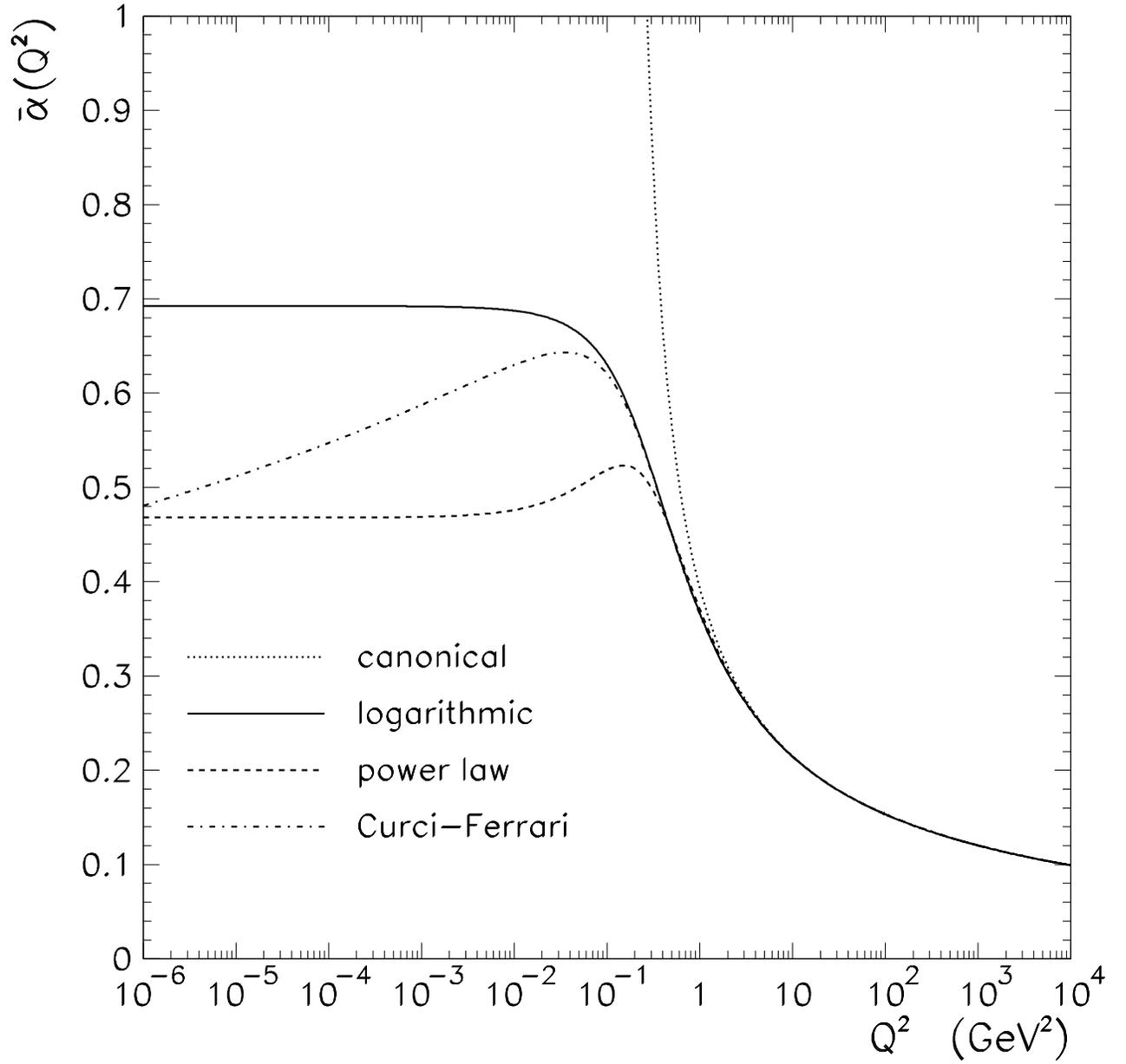}
\caption{The canonical coupling constant and the QCD effective charges at NLO. The logarithmic, power-law and Curci-Ferrari charges are calculated
using $m_{g}=$ 310, 415 and 302 MeV, respectively.}
\label{fig2}
\end{center}
\end{figure*}

\begin{figure*}
\vspace{1.0cm}
\begin{center}
\includegraphics[height=.70\textheight]{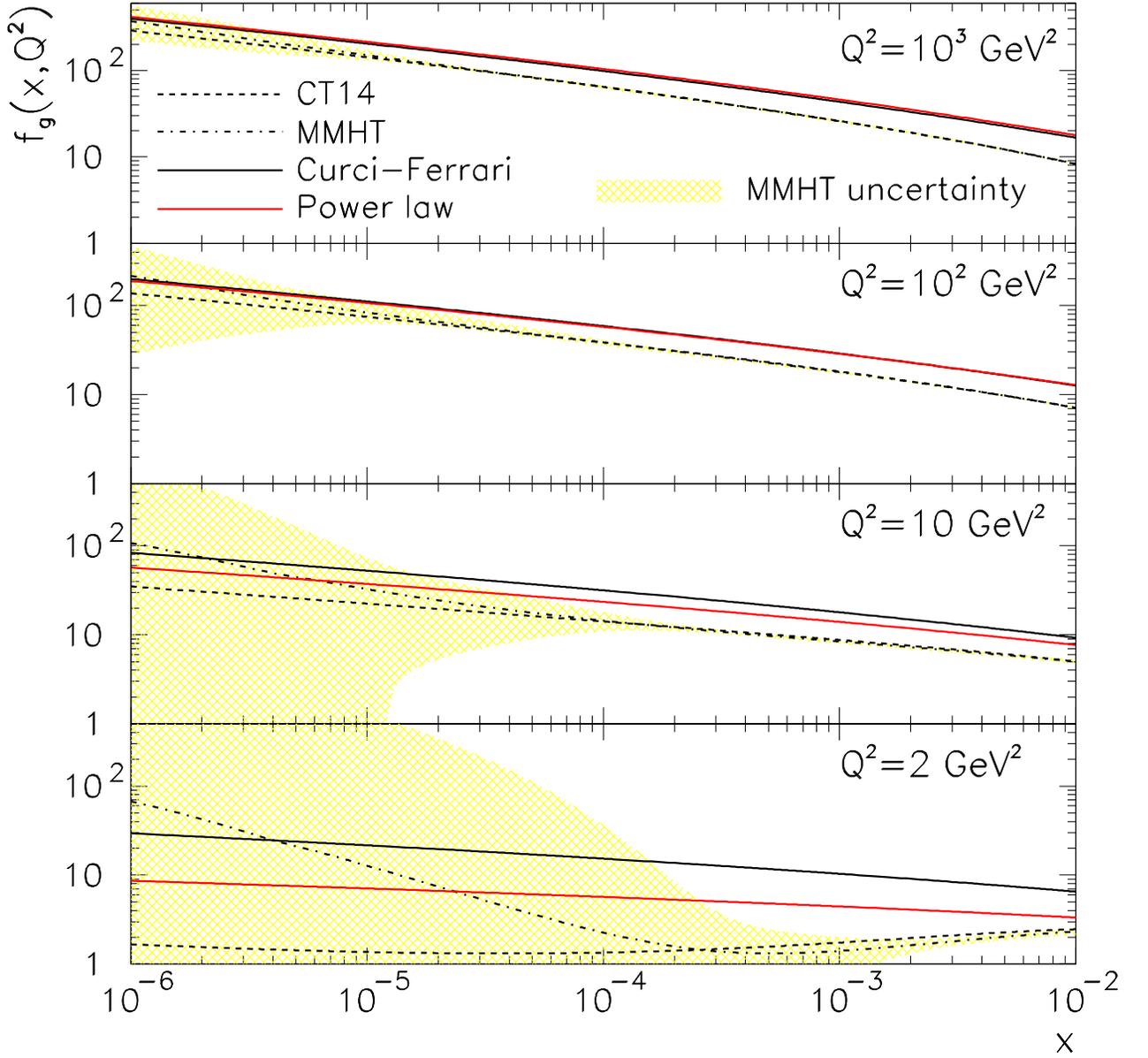}
\caption{The gluon distribution function $f_{g}(x,Q^2)$ at different values of $Q^2$. Our results, calculated using $\bar{\alpha}_{pl}(Q^{2})$ and
$\bar{\alpha}_{CF}(Q^{2})$, are compared to the gluon distribution from the CT14 and MMHT NLO sets.}
\label{fig2}
\end{center}
\end{figure*}

\end{document}